\documentclass[letterpaper]{article} 
\usepackage{aaai23}  
\usepackage{times}  
\usepackage{helvet}  
\usepackage{courier}  
\usepackage[hyphens]{url}  
\usepackage{graphicx} 
\usepackage{natbib}  
\usepackage{caption} 
\usepackage{algorithm}
\usepackage{algorithmic}

\usepackage{newfloat}
\usepackage{listings}
\DeclareCaptionStyle{ruled}{labelfont=normalfont,labelsep=colon,strut=off} 
\floatstyle{ruled}
\newfloat{listing}{tb}{lst}{}
\floatname{listing}{Listing}
\usepackage{amsmath}
\usepackage{amssymb}
\usepackage{mathtools}
\usepackage{pdfpages}
\usepackage{multicol}
\usepackage{booktabs}
\usepackage{comment}
\usepackage{todonotes}
\usepackage{dirtytalk}

\newtheorem{definition}{Definition}

\title{How Do Input Attributes Impact the Privacy Loss in Differential Privacy?}
\author{
    Tamara T. Mueller,\textsuperscript{\rm 1 \rm 2}
    Stefan Kolek, \textsuperscript{\rm 3}
    Friederike Jungmann, \textsuperscript{\rm 2}
    Alexander Ziller, \textsuperscript{\rm 1 \rm 2} \\
    Dmitrii Usynin, \textsuperscript{\rm 1 \rm 2}
    Moritz Knolle, \textsuperscript{\rm 1}
    Daniel Rueckert, \textsuperscript{\rm 1 \rm 4}
    Georgios Kaissis \textsuperscript{\rm 1 \rm 2 \rm 5}
}
\affiliations{
    \textsuperscript{\rm 1} Institute of Artificial Intelligence in Medicine, Technical University of Munich, Germany \\
    \textsuperscript{\rm 2} Institute of Radiology, Technical University of Munich, Germany \\
    \textsuperscript{\rm 3} Department of Mathematics, Ludwig Maximilian University of Munich, Germany \\
    \textsuperscript{\rm 4} Department of Computing, Imperial College London, UK \\
    \textsuperscript{\rm 5} Institute for Machine Learning in Biomedical Imaging, Helmholtz Zentrum Munich \\
}

\usepackage{bibentry}

\begin{document}

\maketitle

\begin{abstract}
Differential privacy (DP) is typically formulated as a worst-case privacy guarantee over all individuals in a database. More recently, extensions to individual subjects or their attributes, have been introduced. Under the individual/per-instance DP interpretation, we study the connection between the per-subject gradient norm in DP neural networks and individual privacy loss and introduce a novel metric termed the \textit{Privacy Loss-Input Susceptibility} (PLIS), which allows one to apportion the subject's privacy loss to their input attributes. We experimentally show how this enables the identification of sensitive attributes and of subjects at high risk of data reconstruction.
\end{abstract}

\section{Introduction and Prior Work}

The handling and processing of sensitive data entails the risk of compromising individual privacy and exposing personal information. Differential Privacy (DP) \cite{dwork2014algorithmic} allows one to offer quantitative privacy guarantees to individuals while allowing to draw conclusions from the dataset as a whole. Under the canonical DP definition, the concept of \say{privacy loss} (PL) refers to a worst-case measure defined over \textit{all} individuals in a dataset. However, such a guarantee may be too coarse for many practical use-cases. Recent works have proposed alternative interpretations of DP, which analyse the PL at the granularity of a single individual (individual or per-instance DP \citep{feldman2021individual, wang2017per}) or even of specific input attributes or features (attribute level DP \citep{asi2019element}). Inherent to all these works is a re-definition of algorithm \textit{sensitivity} from a global measure to a more granular one. In neural network applications, individual sensitivity is linked to the per-subject gradient $L_2$-norm \cite{yu2022per}; since the gradient is considered the sensitive information to-be-published in deep learning, one can think of the per-subject $L_2$-norm as a measure of gradient information content or \say{signal strength}.

Orthogonally to the aforementioned works, recent studies \citep{hannun2021measuring, mo2020layer} have introduced alternative measures of privacy loss, namely the \textit{Fisher Information Loss} (FIL) (going back to an older work of \citet{farokhi2017fisher}) and the Jacobian Sensitivity (\textit{JacSens}). Despite not operating under the DP definition, these measures also exhibit a strong connection to the gradient, as they attempt to quantify the \say{reactivity}\footnote{We eschew the term \say{sensitivity} here to avoid confusion with the notion used in DP.} of individual entries in the gradient to changes in the input space, i.e. the attributes.
In this work, we draw inspiration from both aforementioned lines of work to propose a metric which naturally links the per-sample gradient norm to changes in the input space of the algorithm. Our metric, termed the Privacy Loss-Input Susceptibility (PLIS) represents the Jacobian of the Mahalanobis distance between the per-sample gradient and the origin with respect to the neural network inputs and allows us to apportion the subject's privacy loss to individual input attributes. Our contributions are as follows:    

\begin{itemize}
    \item On the theoretical side, we define the PLIS and motivate it through its strong links to both Rényi and Gaussian DP \citep{mironov2017renyi, dong2021gaussian} as well as to FIL and JacSens.
    \item We evaluate the PLIS in linear models (where a direct comparison to FIL is possible) as well as in convolutional networks and demonstrate that it represents an expressive and easy to compute metric;
    \item We illustrate marked differences in the PLIS distribution of individual attributes between DP-SGD and the non-private setting;
    \item Lastly, we show the link between the PLIS and the detection of out-of-distribution samples as well as to the risk of a successful malicious data reconstruction attack;
\end{itemize}

\section{Theoretical Results}
To motivate PLIS, we recall that --under the Gaussian Mechanism (GM)-- the privacy loss under both Rényi DP (RDP) and Gaussian DP (GDP) depends only on two key quantities: the sensitivity and the noise variance. Concretely, for RDP, we have that a GM $\mathcal{M}$ satisfies $(\alpha, \rho)$-RDP, if, for all pairs of adjacent databases $D$ and $D'$, it holds that:
\begin{equation}
    \mathcal{D}_{\alpha}^{\leftrightarrow}\left(\mathcal{M}(D) \parallel \mathcal{M}(D') \right) \leq \rho,
\end{equation}
where $\mathcal{D}_{\alpha}^{\leftrightarrow}$ is the Rényi divergence of order $\alpha \geq 1$ and the $\leftrightarrow$ superscript denotes $\max \lbrace \mathcal{D}_{\alpha}\left(\mathcal{M}(D) \parallel \mathcal{M}(D') \right), \mathcal{D}_{\alpha}\left(\mathcal{M}(D') \parallel \mathcal{M}(D) \right) \rbrace$. For the GM, the expression admits the following closed form:
\begin{equation}
    \mathcal{D}_{\alpha}\left(\mathcal{M}(D) \parallel \mathcal{M}(D') \right) = \frac{\alpha}{2}\frac{\Delta^2}{\sigma^2},
\end{equation}
where $\Delta$ is the global $L_2$-sensitivity of the query function and $\sigma^2$ the variance of the Gaussian noise. Similarly, for GDP we have that all GMs satisfy $\mu$-GDP with 
\begin{equation}
    \mu= \frac{\Delta}{\sigma}.
\end{equation}
We refer to the original works for details on the definitions, but stress the importance of the term $\frac{\Delta}{\sigma}$ for both. This ratio, which is equivalent to the Mahalanobis distance between the \say{worst-case} subject in the database and the origin, can be interpreted as the signal to noise ratio (SNR) of the mechanism \cite{psi_dp}. For neural network applications and DP-SGD, $\Delta$ is equal to the gradient clipping threshold and represents the upper bound on the permissible $L_2$-norm of the gradient. 
To reformulate the worst-case guarantee over the entire database to an individual/ per-instance privacy loss for DP-SGD, we replace the global $L_2$-sensitivity with the \textit{actual} gradient norm of the subject. We limit our purview to the supervised learning setting. 

\begin{definition}[Individual/Per-instance privacy loss]
Let $\ell(f(x, \theta),y)$ be a loss function of a neural network $f(x, \theta)$, where $(x_i, y_i)$ is a input/label pair belonging to subject $S_i$, and $\boldsymbol{\theta}$ is a vector of learnable parameters. Then the following holds for the individual/per-instance privacy loss: 
\begin{equation}
    \mathbf{PL}(S_i) \propto \frac{\Vert \nabla_{\boldsymbol{\theta}} \ell(x_i | y_i, \boldsymbol{\theta}) \Vert_2^2}{\sigma^2}.
\end{equation}
\end{definition}
We note that the specific assumptions about the database from which $S_i$ is drawn differ between individual and per-instance DP, for which we refer to \citet{feldman2021individual, wang2017per}. We observe that the privacy loss is thus proportional to the Mahalanobis distance of the subject in question's gradient vector to the origin or, in other words, the SNR evaluated at $S_i$. Slightly stretching the definition, we are able to extend it to the non-private setting where no noise is present as:
\begin{equation}
    \mathbf{PL}_{\text{NP}}(S_i) = \Vert \nabla_{\boldsymbol{\theta}} \ell(x_i | y_i, \boldsymbol{\theta}) \Vert_2^2.
\end{equation} 
It is more correct to refer to $\mathbf{PL}_{\text{NP}}(S_i)$ as the \say{gradient signal}, as --formally-- the privacy loss is infinite in the non-private setting, but we will retain the same notation for convenience.
In both cases, the individual/per-instance privacy loss is proportional to the information content/signal of a \textbf{whole} sample in a database. In this work, we are interested in taking this quantification one step further and assess the reactivity of the privacy loss to \textit{each input attribute} of the individual. This quantifies how much impact each input attribute has on the change of the privacy loss of an individual sample and therefore estimates how much privacy is leaked about a specific input attribute of an individual. The natural continuation of the individual/per-instance privacy loss to a per-input-attribute privacy loss is to take the Jacobian of the privacy loss with respect to (WRT) the input attributes. We term this Jacobian matrix the \textit{privacy loss-input susceptibility} (PLIS) matrix:

\begin{definition}[privacy loss-input susceptibility (PLIS)]
Let $\mathbf{PL}(S_i)$ be defined as above. Then, the PLIS is defined as:
\begin{align}
    \mathbf{PLIS}(S_i) &\coloneqq \boldsymbol{J}_{x_i} (\mathbf{PL}(S_i))  \nonumber \\
                &= \boldsymbol{J}_{x_i} \left(\frac{\Vert \nabla_{\boldsymbol{\theta}} \ell(x_i | y_i, \boldsymbol{\theta}) \Vert_2^2}{\sigma^2}\right).
\label{eq:PLIS}
\end{align}
\end{definition}

We observe the following facts about the PLIS definition:
\begin{itemize}
    \item The PLIS matrix has the same dimensionality as the input and assumes high values for the input attributes which contribute to an increase in privacy loss.
    \item The PLIS is easily computable through backpropagation and memory-efficient (the quantity on the RHS of the definition is scalar). This is a notable difference from FIL and JacSens. FIL requires instantiating the Fisher Information Matrix of the inputs WRT the entire gradient vector, whereas JacSens requires materialising a Jacobian matrix of the inputs WRT the gradient vector. Both FIL and JacSens are large (block) matrices and extremely costly to compute (typically not realisable for large neural networks) and have no clear connection to the privacy loss from the DP point-of-view, contrary to PLIS. In our experiments below, we demonstrate that PLIS conveys much of the \textit{same information as FIL} despite its much lower computational cost.
    \item PLIS is related to gradient-based interpretability methods \citep{zhang2021survey} and has a clear operational interpretation. Differently from interpertability techniques, we backpropagate \textit{from the per-sample gradient norm} and not from the (averaged) logits. This is important to retain the individual/per-instance interpretation. Moreover, PLIS also incorporates information of the (private) label instead of showing \textit{which features} are most important for predicting the label.
    \item Similar to all other discussed metrics, PLIS is a private quantity and should not be released and requires full-batch gradient norms (i.e. is incompatible with secret sub-sampling techniques as it needs to assign values to specific individuals). Moreover, it requires a second backpropagation pass and --by extension-- the DP-SGD clipping step to be differentiable (which we realise as described in \citet{hannun2021measuring}).
    \item PLIS corresponds to gradient signal in the non-private setting (alhough we use the abbreviation PLIS for both), allowing us to compare information flow dynamics between private and non-private neural networks.
    \item Like FIL and JacSens\footnote{The authors of this work utilise the Frobenius norm but we will only use the largest singular value $L_2$-norm here.}, the $L_2$-norm of the PLIS matrix (or of a sub-matrix thereof, e.g. a \textit{superpixel} in images) can be used to express the PLIS of the whole individual or a subset of attributes \footnote{However, the $L_2$-norm of the full PLIS matrix is --in practice-- a surrogate of the subject's gradient norm.}.  
\end{itemize}

The similarities between PLIS, FIL and JacSens are not coincidental. They foreshadow a deeper connection between these metrics which can be expressed in a unified way in the language of information geometry (IG). In the language of IG, distributions are points on a \textit{statistical manifold} and the Fisher Information Matrix (FIM) is its metric tensor. It is not coincidental that the Fisher-Rao Distance, i.e. the distance measure on this manifold between two Gaussian distributions with equal variance, is upper-bounded by the Mahalanobis distance between them \citep{pinele2020fisher}. The connection becomes even clearer by expanding the PLIS term as follows:

\begin{align}
    \boldsymbol{J}_{x_i} (\mathbf{PL}(S_i)) &= \boldsymbol{J}_{x_i} \left(\frac{\Vert \nabla_{\boldsymbol{\theta}} \ell(x_i | y_i, \boldsymbol{\theta}) \Vert_2^2}{\sigma^2}\right) = \nonumber \\ = & \frac{2}{\sigma^2}\nabla_{\boldsymbol{\theta}} \ell(x_i|y_i,\boldsymbol{\theta})^T \underline{\boldsymbol{J}_{x_i} \nabla_{\boldsymbol{\theta}}\ell(x_i| y_i,\boldsymbol{\theta})}.
\end{align}
The underlined term on the RHS is exactly equal to JacSens and also occurs in the expression for the sub-matrix of the FIM for individual $S_i$ under the Gaussian mechanism:
\begin{equation}
    \mathbf{FIM}[S_i] = \frac{\boldsymbol{J}_{x_i} \nabla_{\boldsymbol{\theta}}\ell(x_i| y_i,\boldsymbol{\theta})^T\boldsymbol{J}_{x_i} \nabla_{\boldsymbol{\theta}}\ell(x_i| y_i,\boldsymbol{\theta})}{\sigma^2},
\end{equation}
where we abuse notation by expressing an elementwise matrix division by $\sigma^2$ as a fraction. The expressions above also highlight that, whereas DP influences the value of all PLIS and FIL through the noise variance, a bound on the input Jacobian (which PLIS, FIL and JacSens all utilise) is --in general-- much harder to obtain. \citet{hannun2021measuring} utilise the close relationship between the FIM and the Rényi divergence of order $2$ (i.e. $\chi^2$-divergence) to provide a (loose) bound. We however stress that gradient clipping is \textit{not} a valid way to bound the input Jacobian, as it induces a quasi-Lipschitz condition \textit{only} with respect to the weights but not the inputs. Globally Lipschitz objective functions would provide a bound but are not widely used in deep learning. We intend to explore minimax bounds similar to \cite{donoho1990minimax} to answer the question \textit{does a strong bound on PLIS imply a DP bound?} in upcoming work.

\section{Experiments}
\subsection{Relationship between PLIS and FIL}
\begin{figure}[h]
    \centering
    \includegraphics[width=\columnwidth]{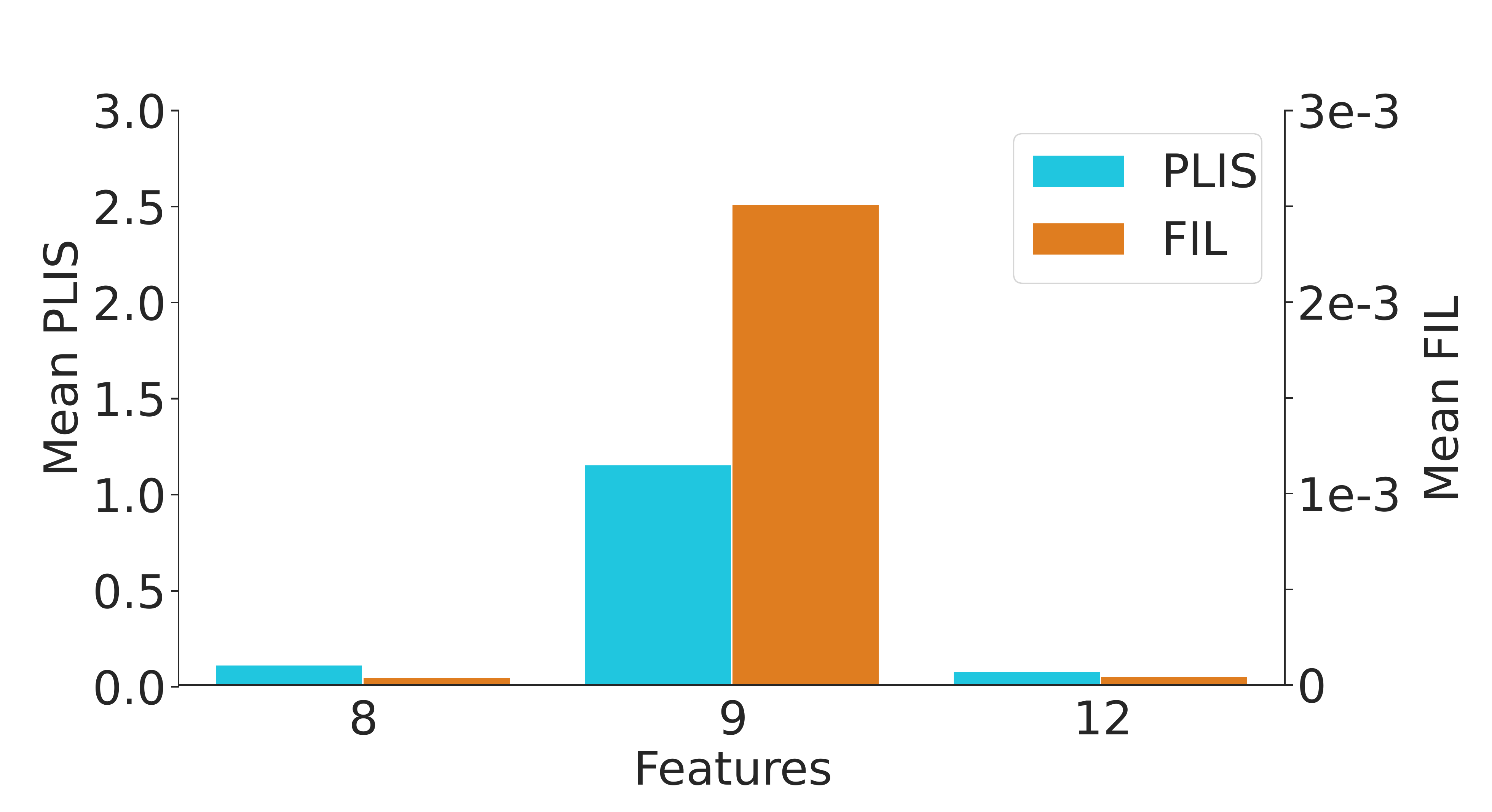}
    \caption{Comparison between FIL and PLIS for one informative ($9$) and two uninformative ($8, 12$) features in a synthetic DP linear regression task showing a strong relationship between the information conveyed by both metrics.}
    \label{fig:fil_PLIS}
\end{figure}

The aforementioned close relationship between PLIS and FIL translates to a strong empirical similarity, which we show in the setting of DP linear regression $(\varepsilon, \delta)=(0.2, 10^{-3})$ on a synthetic dataset. Figure \ref{fig:fil_PLIS} shows the comparison between the average FIL and  PLIS of all data points of the synthetic dataset at convergence for three features. Feature $9$ is an informative feature and exhibits higher privacy losses in both metrics, whereas the uninformative features, exemplarily represented by features $8$ and $12$, have near-zero values for both FIL and PLIS. 

\subsection{PLIS in private and non-private training and for atypical samples}
\begin{figure}[h]
\centering
\includegraphics[width=0.95\columnwidth]{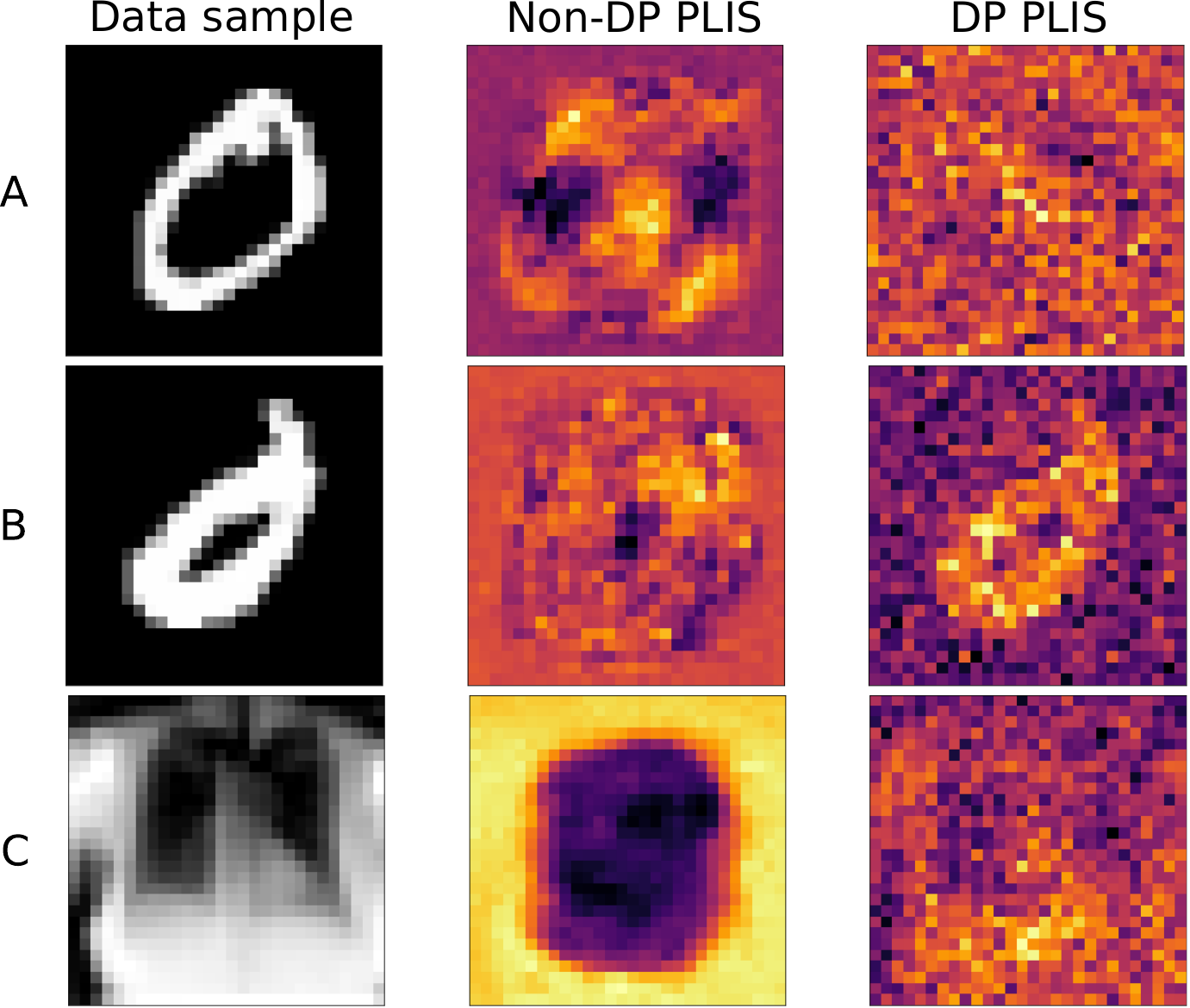} 
\caption{Pixel-wise PLISs for a model trained on MNIST with the input sample on the left and PLIS (or gradient signal) heatmaps for the model with and without DP in the middle and right. \textbf{A} and \textbf{B} show the MNIST samples with the smallest non-zero and highest subject-level PLIS/gradient signal, respectively. \textbf{C} shows an out of distribution sample from MedMNIST.}
\label{fig:mnist}
\end{figure}

Contrary to FIL and JacSens, PLIS can be computed for larger networks. To exemplify properties of PLIS, we begin with a convolutional neural network (CNN) trained with DP-SGD (to $(\varepsilon, \delta)=(1, 10^{-5})$) and without, on the MNIST dataset. Figure \ref{fig:mnist} shows the pixel-wise PLIS of select MNIST samples. We observe that samples with high PLIS are often \textit{atypical} (Figure \ref{fig:mnist}B) or out-of-distribution (Figure \ref{fig:mnist}C). This highlights a noteworthy property of DP: By its very definition, privacy loss depends \textit{only} on the gradient norm without considering \textit{why} a specific norm value occurs. Thus, the fact that a sample has high privacy loss can be due to the fact that it is highly informative or because it is an outlier contributing little to the model. Conversely, samples can be restricted from contributing to training a model (through clipping), which --for highly informative samples-- is a drawback and forms a core motivation behind individual privacy accounting and privacy filters \citep{feldman2021individual}.


Reliable privacy guarantees are especially important in ML settings like medical imaging. We therefore trained a CNN on the chest x-ray subset of MedMNIST with DP-SGD to $(\varepsilon, \delta)=(1, 10^{-5})$, and observed very similar results to the MNIST dataset (Figure \ref{fig:medmnist}). 


\subsection{Reconstruction attacks and PLIS}
In its hypothesis testing interpretation \citep{dong2021gaussian}, DP can best be thought of as a bound on membership inference attack success against the model. However, in many cases, membership privacy can be less important to a subject than a protection against a malicious data \textit{reconstruction attack}. DP is generally very effective against such attacks and formal bounds on reconstruction attack success have been discussed by \citet{guo2022bounding, balle2022reconstructing}. 

\begin{figure}[ht]
\centering
\includegraphics[width=\columnwidth]{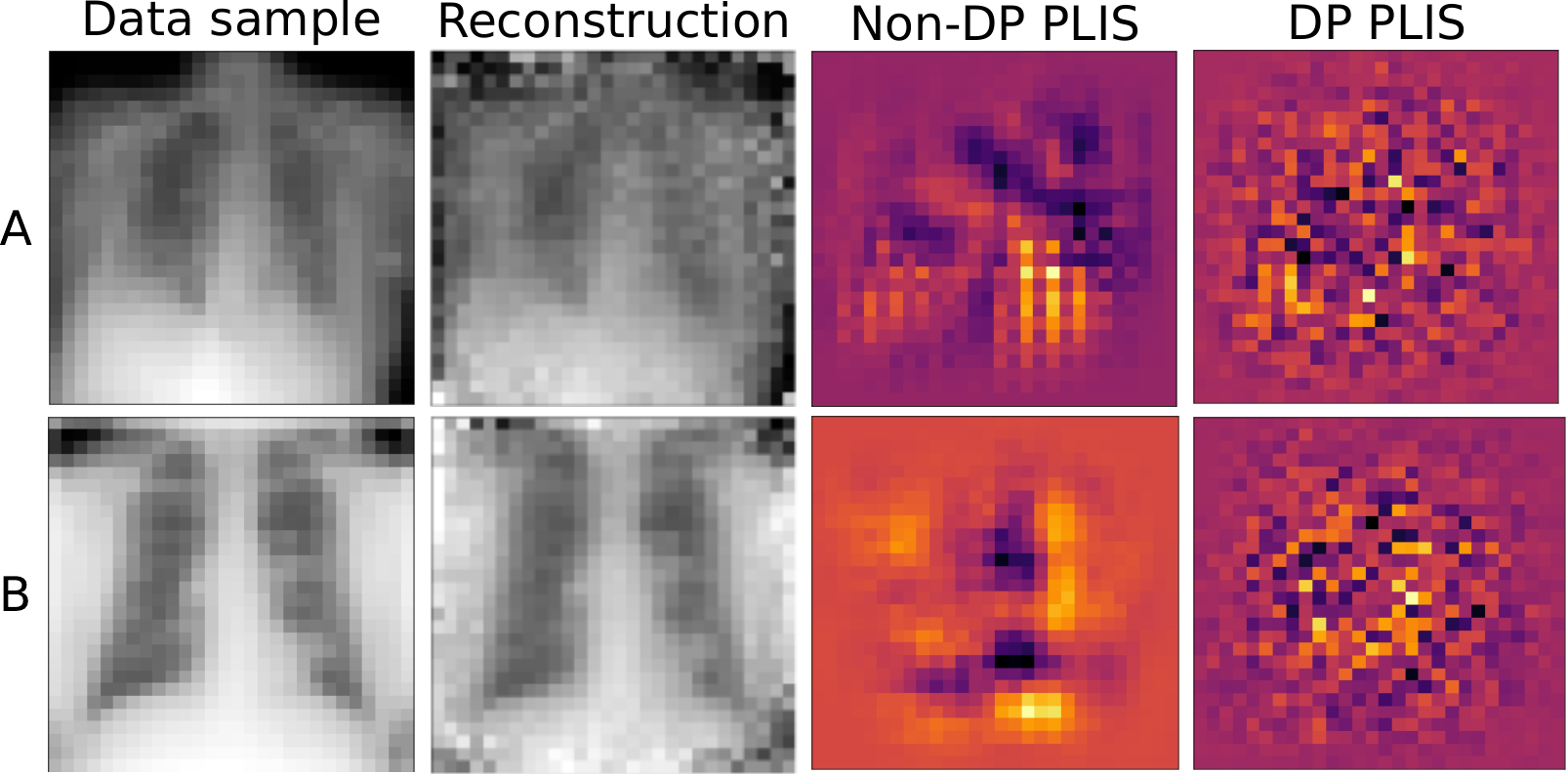} 
\caption{Results of gradient-based reconstruction attack on a data sample with the lowest per-subject PLIS (\textbf{A}) and a data sample with a high subject-level PLIS (\textbf{B}).}
\label{fig:medmnist}
\end{figure}

As mentioned above, we intend to formally investigate the question whether a bound on PLIS implies a DP bound as well as whether it implies a bound on reconstruction attack success, in future work. Here, we empirically evaluate the \textit{reverse} question, that is, whether a higher PLIS value implies a higher success of an actual reconstruction attack against an unprotected model. For this, we execute a gradient-based reconstruction attack using the approach from \citet{geiping2020inverting} on images from the MedMNIST dataset with different subject-level PLISs. The second column of Figure \ref{fig:medmnist} shows the results of the reconstruction attacks on two samples from the MedMNIST dataset, where sub-figure \textbf{A} shows a sample with a low subject-level PLIS and sub-figure \textbf{B} a sample with a high PLIS. Two findings are observed:
\begin{enumerate}
    \item The sample with the higher subject-level PLIS has a higher susceptibility to reconstruction. This also translates to substantially higher objective image quality metrics for the reconstruction from the high-PLIS sample, as shown in Table \ref{tab:scores}.
    \item Comparing the third column of Figure \ref{fig:medmnist} for the low-PLIS sample in sub-figure \textbf{A}, we see that the PLIS is lowest on the top right part of the image (left side of the patient). Correspondingly, the reconstruction of this sample is worst on the top right part of the image. This highlights the added value contributed by the pixel-level PLISs: They pinpoint not only the samples with the highest susceptibility to reconstruction, but also the \textit{areas} in the image which are more likely to end up being reconstructed. 
    \item When trained with DP, no reconstruction was possible, which matches findings in previous literature \cite{balle2022reconstructing}. The rightmost column in Figure \ref{fig:medmnist} shows that the PLIS is essentially randomly distributed over the image space regardless of subject-level PLIS and differently than in the non-DP case, where it is concentrated on salient regions. This fact offers a potential interpretation for the observation that DP is very effective against reconstruction attacks: without a clear concentration in informative attributes, gradient-based reconstruction is nearly impossible.
\end{enumerate}

\begin{table}[h]
    \centering
    \begin{tabular}{lll}
        Metric & Low PLIS (Figure \ref{fig:medmnist} A) & High PLIS (Figure \ref{fig:medmnist} B) \\
        \toprule
        SSIM & 0.8536 & \textbf{0.9912}  \\
        HaarPSI & 0.7323 & \textbf{0.7779} \\
        \bottomrule
    \end{tabular}
    \caption{Similarity scores between the reconstruction and the original image. SSIM: Structural Similarity Index Metric. HaarPSI: Haar Perceptual Similarity Index \citep{reisenhofer2018haar}.}
    \label{tab:scores}
\end{table}

\section{Discussion and Future Work}
In this work, we introduce \textit{privacy loss-input susceptibility} (PLIS), an interpretable measure of attribute-level privacy loss and present our preliminary results by highlighting its theoretical underpinnings, relationships to DP as well as to similar metrics (FIL and JacSens) and benefits, most notably, its deep connection to individual/per-instance DP and its computational efficiency. We show in our experiments, how a high subject-level PLIS can reveal out-of-distribution samples in a dataset and how it is linked to more successful reconstruction attacks.

We consider the following limitations of our work: As a first-order method, PLIS imposes a local linearity assumption and does not satisfy Shapley-like axioms (an extension to satisfy the Shapley axioms is theoretically possible but very computationally expensive). Moreover, as discussed above, bounds on PLIS and their ramifications are still ongoing work that we consider promising future directions. The value of PLIS itself is a sensitive quantity and only useful for \textit{post-hoc} analyses. Furthermore, as noted above, PLIS accounting is incompatible with sub-sampling amplification. 

We view the introduction of a PLIS filter which allows for \textit{ad hoc}/online analyses, as a promising direction for future work. This could allow subjects to allocate a specific privacy budget to subsets of their attributes and aid in the development or improvement of heterogeneous noise mechanisms to selectively protect especially sensitive features.

\bibliography{bibliography}


\end{document}